# Unravelling Airbnb
# Predicting Price for New Listing

**Paridhi Choudhary**
H John Heinz III College
Carnegie Mellon University
Pittsburgh, PA 15213
paridhic@andrew.cmu.edu

**Aniket Jain**
H John Heinz III College
Carnegie Mellon University
Pittsburgh, PA 15213
avjain@andrew.cmu.edu

**Rahul Baijal**
H John Heinz III College
Carnegie Mellon University
Pittsburgh, PA 15213
rbaijal@andrew.cmu.edu

# TABLE OF CONTENTS

# EXECUTIVE SUMMARY

This project analyzes Airbnb listings in the city of San Francisco to better understand how different attributes such as bedrooms, location, house type amongst others can be used to accurately predict the price of a new listing that is optimal in terms of the host's profitability yet affordable to their guests. This model is intended to be helpful to the internal pricing tools that Airbnb provides to its hosts. Furthermore, additional analysis is performed to ascertain the likelihood of a listing's availability for potential guests to consider while making a booking.

The analysis begins with exploring and examining the data to make necessary transformations that can be conducive for a better understanding of the problem at large while helping us make hypotheses. Moving further, machine learning models are built that are intuitive to use to validate the hypotheses on pricing and availability and run experiments in that context to arrive at a viable solution. The project then concludes with a discussion on the business implications, associated risks and future scope.

The presentation of the project can be found at: https://youtu.be/KT0Y1ymTdxM

# PROBLEM STATEMENT

### Motivation
Airbnb is a privately-owned accommodation rental website which allows house owners to rent out their properties to guests looking for a place to stay. Serving as an aggregator for both the house owners and the guests, Airbnb's total valuation exceed 31 Billion dollars in May 2017, with 4.5 million properties listed in 191+ countries.

Airbnb offers complete independence to its hosts to price their properties, with only minimal pointers that allow hosts to compare similar listings in their neighborhood in order to come up with a competitive price. Hosts may incorporate a premium price for any additional amenities they may find necessary. With the number of hosts using Airbnb increasing, coming up with the right price to remain competitive in a host's neighborhood is imperative.

On the flip side, guests using Airbnb are often plagued with non-availability of accommodation due to a variety of reasons. This may include seasonal rush, abrupt booking cancellations, hosts preferences etc.

We propose to analyze Airbnb's publicly available listing information for the past three years to try and validate our hypotheses to alleviate some of Airbnb's issues.

### Hypothesis
1. The hosts on Airbnb experiment and charge an optimal price. So, for a new listing can we analyze similar listings in the past to recommend an optimal the host should charge for the new listing?
   Goal: Hosts get a decent idea how much to charge for a new listing that has no reviews.
2. It is uncertain when a listing is hosted or when it becomes unavailable. Hosts/Guests change plans. So, having the information about availability in the past can we recommend guests with a likelihood of a listing being available?
   Goal: Guests can decide whether to move on or wait for a listing to become available depending on the likelihood.

# EXPLORING THE DATA

### Data Source
The data is scraped from Airbnb's website [1], and hence, is in snapshot format for the years 2015, 2016 and 2017. Data is made available by insideairbnb.com. Data is available for multiple cities all over the world, although our focus for training and testing will be completely restricted to the cities of San Francisco and New York.

The data has been analyzed, cleansed, and aggregated where appropriate to facilitate public discussion.

### Exploratory Data Analysis
The distribution of listing's price is skewed towards the lower price of USD 100 (Figure 1). Although most of the listings fall in this ballpark, roughly 5% of the listings have prices as high as USD 30000.

These outliers may influence the regression lines if a linear regression is fit on this data. We would perform an exercise to remove these outliers.

Another interesting attribute of the listing is the number of bedrooms which varies from 0 to 15 (Figure 2). Although more than 97% of the listings have bedrooms below 4 but these outlier listings can also affect the model trained using this attribute. Such listings are rare, and we do not have a large set of data points to be able to confirm a pattern for these listings. There can be a correlation between the number of bedrooms and high prices or they can be independent too. But having listings with very high price may skew the regression fits which is why we will try to remove these outliers.

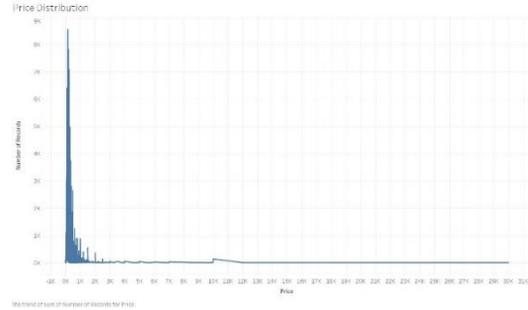

Figure 1: Price Distribution for Listings

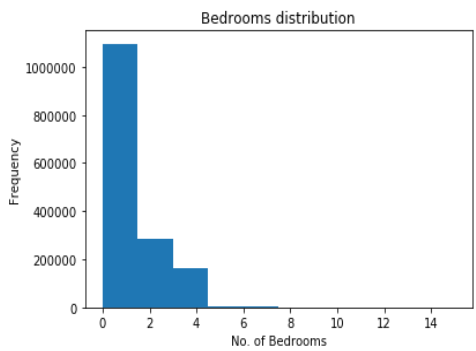

Figure 2: Distribution of # Bedrooms

As people travel over the weekends more where they may want to book an Airbnb and also through personal experience, we had a hypothesis that hosts charge a higher price over weekends than on weekdays. On exploring the data to find if this is true, we found very minimal difference between them and in fact on the contrary, weekend prices were lower than weekday prices (Figure 3).

However, surprised by this finding, we explored the counts of entries of weekend and weekday and found the dataset to be highly imbalanced towards weekday prices. This could be because Airbnb owners use their properties over the weekend themselves and hence less listings are available over the weekend. We balanced the dataset and found the weekend prices to be higher in this case but again the difference was insignificant.

|  | Weekday Median Price (USD) | Weekend Median Price (USD) |
|---|---|---|
| Imbalanced | 186 | 180 |
| Balanced | 164 | 169 |

Figure 3: Weekday Weekend Price Difference

For understanding if there are particular neighborhoods which have higher prices than others, we made a heatmap (Figure 4) of prices over latitude and longitude pairs. The heatmap can be seen which indicates central San Francisco to have a higher price rather than suburban areas but again areas near sea-shores have higher prices. [2]

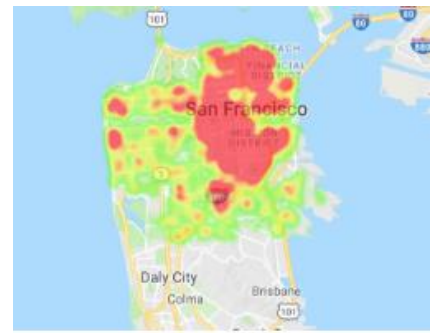

Figure 4: Price Heatmap of San Francisco

Distribution of certain numerical variables like price, cleaning fee, security deposit and charges for extra people were not normal because of the outliers and hence, we log transformed them to be able to use algorithms which assume normal distribution of the predictors such as linear regression. This change in distribution is depicted for the price variable below:

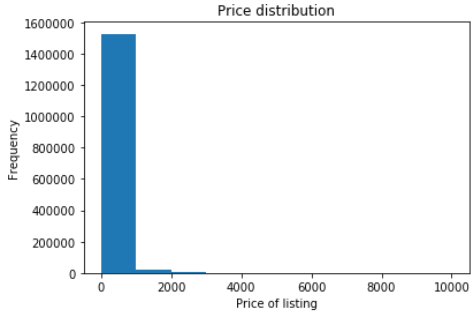

**Figure 5.1: Price Distribution for Listings**

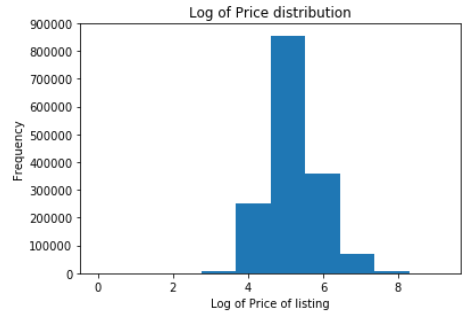

**Figure 5.2: Log of Price Distribution for Listings**

There are no strong correlations (> 0.5 Pearson Correlation). However, Price is slightly positively correlated with the following features of a listing: People Accommodated, Bathrooms, Bedrooms, Cleaning Fee

# METHODOLOGY

# HYPOTHESIS 1: PREDICTING THE OPTIMUM PRICE

We started the prediction of price for a listing by fitting a linear regression and analyzing which predictors are given high importance and check the residuals of the plot.

The RMSE of the prediction came out to be around USD 131 while the mean percentage error is around 28.02%. On a closer look at the distribution of the prediction error (Fig. 6), we can see heavy concentration within an error of USD 30. Checking the percentage of listings in different ranges of errors.

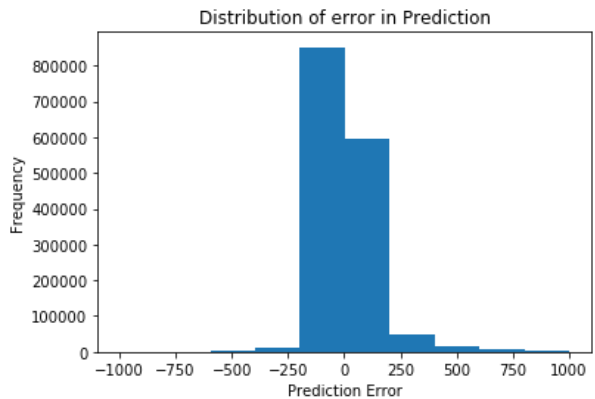

| Prediction Error (USD) | Percentage |
| --- | --- |
| <= 5 | 8% |
| <=10 | 15.93% |
| <=20 | 30.63% |
| <=30 | 44.89% |

**Figure 7: Distribution of Prediction Error**

**Figure 6: Distribution of Prediction Error**

This indicates that almost 50% of the predictions had less than USD 30 error and therefore, some listings are easy to predict while others are little harder. (Fig. 7)

Based on these results and suggestions from our professor we thought we should separate cases of low error and just focus on the low error cases or 'easy' ones.

We need to classify each listing into whether it will be easy to predict or hard to predict before deciding whether we will be predicting it. To achieve this, a random forest classifier was trained [3]. The features that were deemed important according to this classifier are given in the figure on the right. The order indicates decreasing importance of predictors (Figure 8).

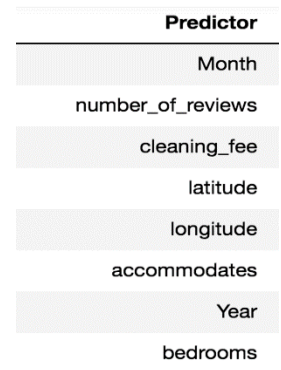

**Figure 8: Important features**

On checking how the model differentiates between easy and hard to predict samples, we could see some predictors as good differentiators like below. Cleaning fee if less than 4 is less likely to be easy to predict (Fig 9.1). However, some features though deemed important did not provide an obvious differentiating line like below. Every month has similar frequency of easy and hard samples (Fig 9.2).

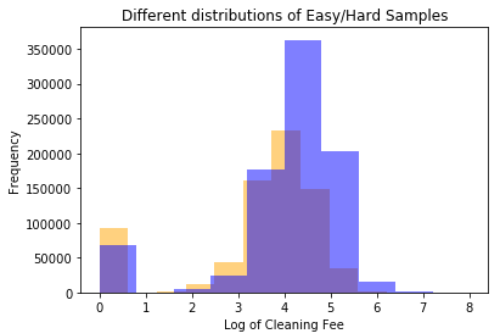
Figure 9.1: Easy vs Hard Cleaning Fee distribution

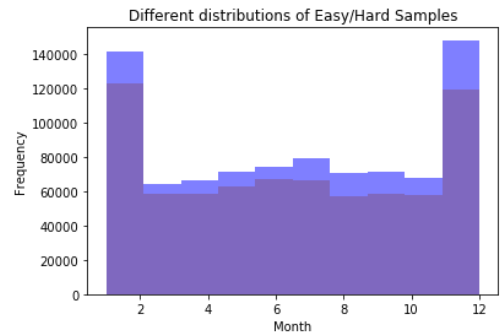
Figure 9.2: Easy vs Hard Month distribution

The model that was trained for easy cases had encouragingly low training error but discouragingly high test error. This implied overfitting in the training which was then followed by two suggestions:

1. Apply Cross Validation: Train and test against different folds to identify more robust set of easy. Then train the classifier to test against the unseen cases.
2. Upsample and Downsample: Until this point we had some listings for which we have a lot of price instances (as high as 4000) and others which have only 1 instance. This could lead to bias the model towards the highly frequent listings thinking there is a pattern underlying. Thus, we were advised to balance the dataset by upsampling the listings which were rare and downsampling the ones that were frequent.

### Methodology to upsample and downsample

The idea is to have equal number of instances for each listing so that they have equal weights in the model. We had 7000 unique listings in our dataset. We decide to use 100 as the optimal number of instances required for each listing as that we will give us a dataset of around 700,000 samples which is optimal to work with.

We needed to perform both upsampling and downsampling.

**Upsampling**: For upsampling, we took a listings unique entry in the dataset and repeated them to reach the length of 100.

**Downsampling**: For downsampling, we found the median of the dataset which gives a row in the dataset which is the center of the dataset looking at all columns together. Then we find the Euclidean distance of each row from this median row [4]. This is followed by sorting the rows according to their distance from the median and take the top 100 values. We used this approach to find most diverse combinations of features so that we cover a larger space and ignore samples which are very similar to each other in feature values and prices.

After balancing the dataset, we wanted to tune the hyper parameters for the random forest regressor. We used sklearn's RandomizedSearchCV [5] and ran 100 iterations for a set of parameters. The grid of parameters used was:

**N_estimators** - [100 - 200]

**Max_features** - ['auto', 'sqrt']

**max_depth** - [10 - 20, None]

**min_samples_split** - [2, 5, 10]

**min_samples_leaf** - [1, 2, 4]

**bootstrap** - [True, False]

The 100 iterations were run using 10-fold cross validation to find best parameters which can be used to train the dataset.

Top 3 ranked parameters were:

1. Model with rank: 1

Mean validation score: 0.980 (std: 0.001)

Parameters: {'n_estimators': 131, 'min_samples_split': 5, 'min_samples_leaf': 2, 'max_features': 'auto', 'max_depth': None, 'bootstrap': True}

2. Model with rank: 2

Mean validation score: 0.979 (std: 0.001)

Parameters: {'n_estimators': 142, 'min_samples_split': 2, 'min_samples_leaf': 1, 'max_features': 'sqrt', 'max_depth': None, 'bootstrap': True}

3. Model with rank: 3

Mean validation score: 0.978 (std: 0.001)

Parameters: {'n_estimators': 200, 'min_samples_split': 2, 'min_samples_leaf': 2, 'max_features': 'sqrt', 'max_depth': None, 'bootstrap': False}

The RMSE of the prediction was USD 28 with a mean percentage error of 0.04%.

The figure below shows the distribution of the prediction error of samples which had lower than USD 5. 99% listings had prediction error less than USD 5 (Fig 10). We also tried to check the performance of the model on increasing the number of trees used to train the model. The training and test error are averaged over 10 folds of 10-Fold CV (Fig 11).

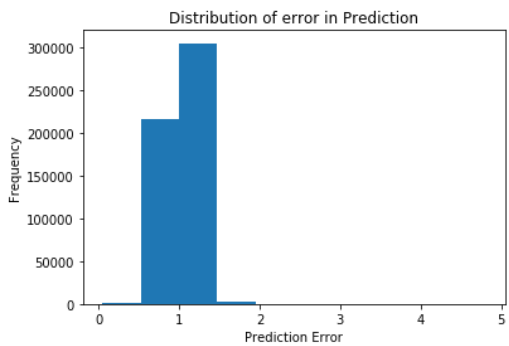
**Figure 10: Distribution of Prediction Error**

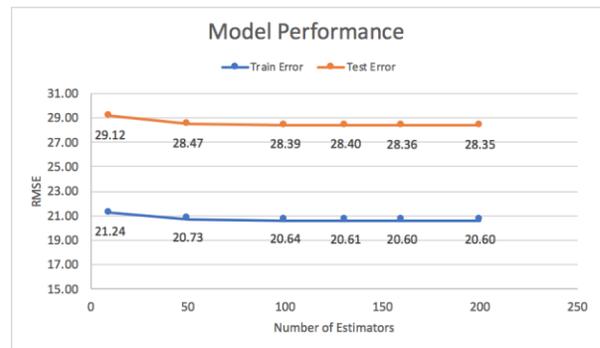
**Figure 11: Random Forest Regressor performance with change in number of trees**

# HYPOTHESIS 2: LIKELIHOOD OF AVAILABILITY

**Segregating the listings into high/low availability – Clustering + Naïve Bayes**

Our motivation for applying clustering to our data was to ascertain the structure of given availabilities (#days/year that the property is available to be rented out) of property listings. The data presents 4 types of availabilities namely 30 days/year, 60 days/year, 90 days/year and 365 days/year. Our aim is to group each type into clusters of low and high availability (Fig 12). (Note: X-Axis is normalized over #days)

For 30 days/year, clusters were made of equal size possibly because of the small range of days. For 30 days/year and 90 days/year, clusters were better defined.

Further, we wished to classify each listing into high and low availability. Taking 'zip code' and 'room type' as features, we trained a multinomial Naïve Bayes model for classification.

This model was 64.1% accurate, whereas the default model's accuracy was 62.8%.

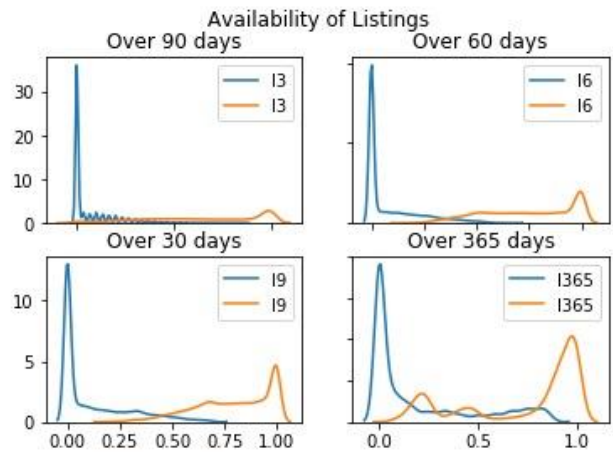

**Figure 12: Resulting Distribution of Clustering**

Future Recommendations:

1. Include more features to get sharper classification of high or low availability.
2. Obtain an optimum threshold over the number of days to classify new listings as high or low available. This gives way to answer the question: If I host a new listing, how busy will it's booking will be in the next 30/60/90 days.

# RESULTS AND CONCLUSION

The difference in the performance of same RandomForestRegressor on the balanced and imbalanced dataset confirms the hypothesis that high frequency of some samples was skewing the predictions and overfitting for them. Therefore, we decided to go ahead with the balanced dataset and RandomForestRegressor.

Considering the initial hypothesis of whether we will be able to predict price of a new listing based upon its features, we would say that we will be able to predict the prices within USD 29. This is aligned with the project goals because the user can use our initial price to start with and then make necessary changes based on interaction with the customers.

# BUSINESS IMPLICATION AND RISK MITIGATION

Using just publicly available data about the Airbnb listings we have looked at approaches building upon predicting an optimal price for a new listing. Hosts can make use of historical data to get an idea about how much other hosts are charging for similar listings. We believe this adds practical value to the client as they can not only get an idea of the prices they should charge but also gives a chance to mark up/down the prices.

Airbnb might have price tips based on multiple sources of information. Although, we or the hosts do not know if they bias these prices based on personal interests. An example can be that Airbnb wants their listing prices varied so more customers are converted. So, the additional value this tool adds is an unbiased estimator.

Considering just for San Francisco, the neighborhoods play a less significant role in setting prices. The hosts should make the best of the proposed features such as time seasonality, listing characteristics, analysis of current prices for similar listings on Airbnb and finally make use of this tool to get the optimum price they should charge for their new listing.

The assumption is that the publicly available prices are "successful" prices for hosts as they experiment over time and better their prices against customer affordability and maximize income. The risk is that this assumption can be wrong and the booking history for guests could provide more useful information as "successful" prices. The way to mitigate this risk is to do validate the similarity between the prices guests are charged and the prices the hosts are listing.

## FUTURE SCOPE

We have features that were not used in this exercise because of time limitations. These are majorly text based features and would require natural language processing to convert into features. In continuation of the work, these features can be used:
1. Amenities (Text)
2. Street
3. Description (Text)
4. Reviews (Text)
5. Interactions between owner and guest (Text)
6. Ratings
7. Reviews_per_month

Also, we identified that for each neighborhood, there are centres where the price is higher but as you move away from the centre, the price reduces. It is visible in the heatmap shown previously.

## REFERENCES

1. http://insideairbnb.com/get-the-data.html
2. GMAPS tutorial: https://github.com/pbugnion/gmaps
3. Random Forest Classifier: http://scikit-learn.org/stable/modules/generated/sklearn.ensemble.RandomForestClassifier.html
4. Dataframe Median: https://pandas.pydata.org/pandas-docs/stable/generated/pandas.DataFrame.median.html
5. RandomizedSearchCV http://scikit-learn.org/stable/modules/generated/sklearn.model_selection.RandomizedSearchCV.html